\documentclass[aps,prl,showpacs,groupedaddress,twocolumn]{revtex4}
\usepackage{amsfonts}
\usepackage{amsmath}
\usepackage{mathrsfs}
\usepackage{amssymb}
\usepackage{graphicx}
\usepackage{dcolumn}
\usepackage{bm}
\usepackage{color}
\usepackage{endnotes}
\begin{document}
\title{Quantum optomechanics of a Bose-Einstein Antiferromagnet}
\author{H. Jing, D. S. Goldbaum, L. Buchmann, and P. Meystre}
\affiliation{B2 Institute, Department of Physics and College of
Optical Sciences, The University of Arizona, Tucson, Arizona 85721}
\date{\today}
\newcommand{\dhf}{\Delta_{\rm hf}}
\newcommand{\mub}{\mu_{\rm B}}
\newcommand{\omth}{\omega_\theta}
\newcommand{\dth}{D_{\theta}}
\newcommand{\hr}{H_{\rm R}}
\newcommand{\hc}{H_{\rm C}}
\newcommand{\hl}{H_{\rm L}}
\newcommand{\hd}{H_{\dth}}
\newcommand{\hg}{H_{\gamma}}
\newcommand{\kl}{\kappa_{\rm L}}
\newcommand{\xith}{\xi_{\theta}}
\newcommand{\oml}{\omega_{\rm L}}
\newcommand{\omc}{\omega_{\rm C}}
\newcommand{\oma}{\omega_{\rm a}}
\newcommand{\hopt}{H_{\rm OM}}
\newcommand{\ain}{a^{\rm in}}
\newcommand{\aind}{a^{{\rm in},\dag}}
\newcommand{\epsin}{\varepsilon_{\theta}^{\rm in}}
\newcommand{\depsin}{\delta \epsin}
\newcommand{\psip}{\psi_+}
\newcommand{\psipd}{\psip^\dagger}
\newcommand{\psin}{\psi_-}
\newcommand{\psind}{\psin^\dagger}
\newcommand{\psiz}{\psi_0}
\newcommand{\psizd}{\psiz^\dagger}

\begin{abstract}

We investigate the cavity optomechanical properties of an
antiferromagnetic Bose-Einstein condensate, where the role of the
mechanical element is played by spin-wave excitations. We show how
this system can be described by a single rotor that can be prepared
deep in the quantum regime under realizable experimental conditions.
This system provides a bottom-up realization of dispersive
rotational optomechanics, and opens the door to the direct
observation of quantum spin fluctuations.

\end{abstract}
\pacs{42.50.Pq, 03.75.Mn} \maketitle

Cavity optomechanics (COM) has witnessed considerable advances in
recent years, promising significant
contributions in several topics of broad interest including the
production and manipulation of macroscopic quantum states,
the investigation of the crossover between quantum and classical
regimes in systems of increasing size, and the realization of a wide
range of applications from high-precision metrology to quantum
information science and motion-mediated nonlinear photonics
\cite{rev,1photon,disper}.


The first such systems featured optomechanical couplings with a
linear dependence on the displacement of a vibrating mechanical
element such as a mirror~\cite{rev}.
More recently, systems featuring quadratic optomechanical coupling have also been realized~\cite{disper}.
This advance was quite important
since, in principle, quadratic optomechanical coupling allows the
non-destructive measurement of the phonon number state of the
oscillator, as well as observation of the discrete quantum jumps
that accompany changes in phonon number \cite{disper,disperbec}.
In contrast, these measurements are impossible with only
linear optomechanical coupling.

A parallel development in COM was the realization of COM
systems formed by ultracold atomic gases trapped inside driven
high-Q resonators~\cite{optobec}, the role of the macroscopic
mechanical element now being played by a collective excitation of
the atoms. These systems can be thought of as ``bottom-up"
realizations of COM, since in practice one can readily prepare a
degenerate atomic gas in its quantum ground state, and thus the
motion of the fictitious mechanical element is initially in the
quantum regime \cite{optobec}. This is in contrast to the more
conventional ``top-down" COM systems described above where despite
considerable progress  it remains in general an ongoing challenge to
cool the optomechanically-coupled mechanical mode into the quantum
regime.

This Letter considers a bottom-up situation where the role of the
mechanical element is played by spin-wave excitations of a
macroscopic mode of an antiferromagnetic Bose-Einstein condensate
(BEC), whose dynamics can be mapped to a quantum rotor model
\cite{rotor1,rotor2,rotor3,rotor4}. Although there have been multiple
investigations of the cavity dynamics of a spin ensemble
\cite{cavspin,Kurn,zhang,optody,CavityMRI}, as well as an atom-chip-based
realization of quadratic-coupling COM~\cite{disperbec}, this is to our knowledge the first study of bottom-up COM where the atomic gas has spin degrees
of freedom and the cavity confinement leads to the ability to
resolve the quantum regime of an antiferromagnetic spin gas.

Especially pertinent to the present study is Ref.~\cite{Kurn}, which provides the first link that we know of between the fields of spin systems and COM control in the literature. That paper studies the dynamics of a large quantum spin in an optical resonator and finds spin bistability and squeezing of light by establishing an analogy with a harmonic oscillator coupled {\it linearly} to a single cavity mode. In contrast, the present Letter describes an experimentally realizable process for the dramatic cooling of a collective spin-wave excitation mode by establishing an analogy with a torsional oscillator coupled {\it quadratically} to a single cavity mode. As already mentioned the form of the optomechanical coupling is crucial in determining which phenomena can be observed in such a system.

We consider an atomic spin-1 Bose condensate characterized by
antiferromagnetic spin-exchange collisions, and described by the
Hamiltonian ($\hbar =1$)
\begin{eqnarray}
H_{\rm R}=\frac{c_2}{2N}F^2-q\psi^\dag_0\psi_0 \, .
\label{eq1}
\end{eqnarray}
Here $c_2$ is the antiferromagnetic spin coupling, $N$ is the total particle number, and $F=\psi_{\rm i}^\dag F_{\rm
ij}\psi_{ \rm j}$ is the total spin operator, where $F_{\rm ij}$ are
the spin-1 matrices and $\psi_{\rm i}~({\rm i}=\pm,~0)$ are the
bosonic annihilation operators for each spin component. The
additional coupling, $q=(\mub B)^2/(4\dhf)$, is the quadratic Zeeman
shift, with external magnetic field $B$ and hyperfine splitting
$\dhf$.

Recently, it has been shown that there is an exact mapping of the Hamiltonian~\eqref{eq1} to a Hamiltonian describing a single quantum rotor~\cite{rotor3, Note1},
\begin{eqnarray}
H_{\rm R}&=&\frac{L^2}{2I}+V(\theta),\nonumber\\ V(\theta)&=&q(N+3/2)\sin^2\theta+\frac{q^2N}{8c_2}\sin^2(2\theta),
\label{qrmodel}
\end{eqnarray}
where the conjugate operators $\theta$ and $L$ are the rotor's effective angular displacement and angular momentum, respectively, and the effective moment-of-inertia is $I=N/c_2$. An effective position space basis is specified by defining the eigenstates of the Cartesian components $|X\rangle=(|+\rangle-|-\rangle)/\sqrt{2}$, $|Y\rangle=(|+\rangle+|-\rangle)/i \sqrt{2}$, and $|Z\rangle=|0\rangle$, where $\{ \lvert  \pm \rangle, \lvert 0 \rangle \}$ is the single-particle spin eigenbasis. 

We focus on the weak magnetic field limit $q \ll c_2$ where the dynamics of the quantum rotor are dominated by the first term of $V(\theta)$, and are thus localized around the poles of the Bloch sphere. By expanding $V(\theta)$ to quadratic order about the minimum $\theta=0$ we find that the quantum rotor model of Eq.~\eqref{qrmodel} is further reduced to one that describes a one-dimensional harmonic torsional oscillator characterized by $\theta$ and $L_z$, where $[\theta, L_z] = i$ ~\cite{rotor2,rotor3}, and $\theta$ is the polar angle of the total spin state.
In that limit the approximate potential is
\begin{eqnarray}
V(\theta) \simeq \frac{1}{2}I\omega^2_\theta\theta^2,~~ \omth^2=2 q c_2 \left[1+\frac{3}{2} \frac{1}{N} + \frac{q}{c_2} \right],
\end{eqnarray}
for which the ground state wave-function is~\cite{rotor3}
\begin{align}
\Psi_0(\theta)=\sqrt{\frac{1}{\pi\bar{\theta}^2}}\exp\left(-\frac{\theta^2}{2\bar{\theta}^2}\right),~~\bar{\theta}=\sqrt{\frac{c_2}{2qN^2}} \, .
\end{align}
One finds however that the harmonic torsional oscillator description is only valid when $\bar{\theta} \ll 1$. Combining this requirement with the initial weak-field condition, the parameter region where harmonic approximation of the quantum rotor is valid is
\begin{align}
1\ll c_2/q \ll 2N^2 \, .
\label{condition}
\end{align}
When this condition is met, any finite field will localize the state of the spinor condensate about $\theta=0$ regardless of its initial state~\cite{rotor3,ho}.
In this so-called ``Josephson regime''~\cite{rotor4} the rotor model can be used to
predict several purely quantum mechanical phenomena such as
small-number effects ($c_2/q=20,~N=200$ or $\bar{\theta}\simeq
0.02$) \cite{rotor2}, observation of spin collapse and revival
($\bar{\theta}=0.1$)~\cite{rotor3}, and nematic-mixing
dynamics~\cite{pu}. These effects are difficult to observe
experimentally because for coupling strengths characteristic of
antiferromagnetic BEC, mean-field effects will dominate unless one uses relatively small condensates, $N\sim$ tens to hundreds of atoms~\cite{rotor2}. This requirement is
in conflict with the requirement of relatively large condensates
needed to characterize them through the usual probe of absorption
imaging. As a result most studies of spinor BECs have
focused on mean-field physics \cite{ho}.

However, when an antiferromagnetic BEC is trapped in a driven
high-$Q$ optical cavity, the atom-photon coupling is greatly
enhanced, and can thus be used as a sensitive experimental probe for
the BEC. It follows that, under presently available experimental
conditions (see, for example, Ref.~\cite{CavityMRI}), one can measure the quantum dynamics of small
condensates, and as we will show, even observe the quantum regime in
large condensates. More specifically, such cavity confinement allows
one to enter the quantum regime where an antiferromagnetic spin-1
BEC is described by the quantum rotor model.

When trapped inside a unidirectional ring cavity the antiferromagnetic BEC is described by the Hamiltonian
\begin{equation}
H=H_{\rm R}+H_{\rm C}+ H_{\rm L}+H_{\dth}+H_{\gamma},
\end{equation}
where $\hr=L_z^2/(2 I)+(1/2)I \omth^2 \theta^2$ is the quantum rotor
Hamiltonian, $\hc$ describes the intracavity field and its
interaction with the rotor, $\hl= -i\kappa_L(a-a^\dag)$ describes
the laser pumping with amplitude $\kl$, and $\hd$ and $\hg$ account
for mechanical and  input noise, respectively.

The effective optomechanical coupling arises within $\hc$. In the rotating frame of the pump laser
\begin{align}
\hc=&(\omc-\omega_L)a^\dag a +U_0(\psi_+^\dag \psi_+
+\psi_-^\dag
\psi_-)a^\dag a \, ,
\label{hceq}
\end{align}
where $\oml$ and $\omc$ are the frequencies of the pump laser and
the empty cavity resonance, respectively. The second term of
Eq.~\eqref{hceq} describes the far off-resonant coupling of the
cavity field to the BEC, where $U_0=g^2/(\oml-\oma)$ is the
single-photon light shift with dipole coupling $g$ and atomic
transition frequency $\oma$. Assuming that the cavity field is
$\pi$-polarized, the atoms in the $F=1$ ground-state manifold couple
to the $F^\prime=1$ excited-state manifold. The selection rules for
the corresponding transitions are $\Delta m_{ F}=0$, with the
exception that the $\lvert F=1, m_{ F} =0 \rangle \rightarrow \lvert
F^\prime = 1, m_{ F^\prime} = 0 \rangle$ transition is forbidden,
which means that atoms in the spin-0 state do not couple to the
cavity field~\cite{zhang}.

With the normalization condition $\psipd \psip + \psind \psin + \psizd \psiz = N$, the mapping $-q \psizd \psiz \rightarrow V(\theta)$, and in the harmonic approximation $\hc$ can be reexpressed as
\begin{equation}
\hc \simeq (- \Delta + \xith \theta^2) a^\dagger a \, ,
\end{equation}
where $\Delta= \oml-(\omc + U_0 N)$ is the static part of the pump-cavity detuning. The dynamic part of the pump-cavity detuning is readily identified as the quadratic optomechanical coupling term $\hopt=\xith \theta^2 a^\dag a$, with
\begin{align}
\xi_\theta = U_0 N \left[1+\frac{3}{2} \frac{1}{N} + \frac{q}{c_2} \right]\, .
\end{align}

The most salient feature of $\hopt$ is the quadratic dependence on
$\theta$, which, in principle, allows one to perform quantum
non-demolition measurements of the rotational energy of the
oscillator as well as to observe the discrete quantum jumps that
accompany a change in roton number \cite{disper,disperbec}, thereby
providing a unique probe of the quantum state of an
antiferromagmetic BEC. We remark that, as in other studies of
systems featuring quadratic optomechanical coupling
\cite{disper,disperbec,nunnen}, one can adjust the system
parameters, in this case tune the magnetic field, to realize higher
order couplings. For example
\begin{equation}
H_{\theta^4}=H_{\theta^2}-\beta\theta^4,~~\beta\simeq (q-U_0a^\dag
a)N/3.
\end{equation}
where $H_{\theta^2}$ is the quadratic coupling Hamiltonian discussed above.

We now turn to the system dynamics of $H_{\theta^2}$, which we study by considering the corresponding Heisenberg-Langevin equations
\begin{align}
\dot{\theta}=&L_z/I, ~~~~
\dot{L_z}=-I\omega^2_\theta\theta-2\xi_\theta a^\dag
a\theta-\frac{D_\theta}{I}L_z+\epsin,\nonumber\\
\dot{a}=&-i(-\Delta+\xi_\theta\theta^2)a-\gamma
a+\kappa_L+\sqrt{2\gamma} \ain.
\label{EOM}
\end{align}
Here $\gamma$ is the cavity damping rate; $\ain$ is the operator describing the input noise, characterized by a zero mean and Markovian correlations $\langle \ain(t) \aind(t')\rangle=\delta(t-t')$,  and $\epsin$ accounts for noise from a thermal reservoir and/or incoherent atomic scattering, with $\langle \epsin \rangle=0$ and
\begin{align}
\langle \epsin(t)\epsin &(t') \rangle=\nonumber\\
&D_\theta\int^{+\infty}_{-\infty}\frac{d\omega}{2\pi}e^{-i\omega(t-t')}\omega \left[1+\coth(\frac{\omega}{2k_BT}) \right] \, , \nonumber
\end{align}
where $\dth$ is the intrinsic damping constant for the quantum rotor and $k_{\rm B}$ is Boltzmann's constant.

In the mean-field approximation, the steady-state solutions of the Heisenberg-Langevin equations are
\begin{eqnarray}
\theta_{\rm s}=L_{z,\rm s}=0,~~a_{\rm s}=\frac{\kappa_L}{\gamma-i\Delta}.
\end{eqnarray}
An important feature of these solutions is the lack of
multistability, a common hallmark of many cavity optodynamical
systems, including both the linear optorotational coupling case
\cite{rotate} and the general cavity-spin coupling case
\cite{cavspin,zhang,Kurn,optody}. The lack of a multistable region
plays a pivotal role in the enhanced cooling and trapping of the
effective quantum rotor: it allows one to achieve stronger cooling
and trapping by increasing the pump laser power without having to
worry about instabilities arising due to the emergence of additional
stable states. Of course, when the harmonic motion condition of
Eq.~\eqref{condition} is severely violated, such as in the limit of
high magnetic fields $(q> c_2)$ treated in Ref.~\cite{zhang}, strong
bistability tends to occur.

Similarly to the case of a linearly coupled rotating
mirror~\cite{rotate}, the {\it quadratic} optomechanical coupling of
the quantum rotor results in an increased effective trapping
frequency. We show this by replacing each system operator with its
first-order fluctuation expansion $\hat{\mathcal{O}}(t)\equiv
\mathcal{O}_{\rm s}+\delta\hat{\mathcal{O}}(t)$, writing the cavity
field fluctuation in terms of its quadrature operators $\delta
X_1=(\delta a+\delta a^\dag)/\sqrt2$, $\delta X_2=(\delta a-\delta
a^\dag)/i\sqrt2$, and then linearizing the Heisenberg-Langevin
equations with respect to fluctuations to obtain
\begin{align}
\dot v(t)&=Rv(t)+\Lambda(t),~~~~
v(t)=(\delta\theta, \delta L_z, \delta X_1, \delta X_2)^T,\nonumber \\
\Lambda(t)&=(0, \epsin, \sqrt{2\gamma}  X_1^{\rm in},
\sqrt{2\gamma} X_2^{\rm in}), \end{align}
 where
\begin{eqnarray}
R=\left(
\begin{array}{cccc}
0 & 1/I& 0 & 0 \\
-I\omega^2_\theta-2\xi_\theta|a_{\rm s}|^2 & -D_\theta/I & 0 & 0 \\
0 & 0 & -\gamma  & -\Delta \\
0 & 0 & \Delta & -\gamma
\end{array}
\right)\, ,
\end{eqnarray}
and $X^{\rm in}_1=(\ain +\aind )/\sqrt2$ and $X^{\rm in}_2=(\ain-\aind)/i \sqrt2$ are the quadrature operators for the input-field fluctuations. It is important to note that a steady state solution is only stable if the real part of each of the corresponding eigenvalues of $R$ is non-positive, which can be easily confirmed by using the Routh-Hurwitz criterion \cite{RZ}.

After Fourier transform and subsequent algebraic manipulation one solves for the first-order fluctuations
\begin{align}
\delta\theta[\omega]=&\frac{\epsin}{I(\omega_\theta^2-\omega^2)+2\xi_\theta|a_{\rm s}|^2-i\omega
D_\theta},
\label{the}
\\
\delta X_{1,2}[\omega]=&\frac{\sqrt{2\gamma}\{(\gamma-i\omega)
X_{1,2}^{\rm in}[\omega]\pm {\Delta}
X_{2,1}^{\rm in}[\omega]\}}{{\Delta}^2+(\gamma-i\omega)^2}.
\end{align}
Equation~\eqref{the} is in the form of a linear response
$$
\delta\theta[\omega]=\chi(\omega) \, \epsin[\omega],
$$
where the noise operator $\epsin$ plays the role of the external perturbation, and with
\begin{equation}
\chi^{-1}[\omega]=I(\omega^2_{\rm eff}-\omega^2)-i\Gamma_{\mathrm{eff}}\omega,
\end{equation}
where $\chi(\omega)$ is the susceptibility of a fictitious mechanical rotor characterized by the damping factor $\Gamma_{\rm eff} = \dth$ and the resonant frequency
$\omega_{\mathrm{eff}}=\eta \, \omega_\theta$,
where we have defined the enhancement factor
\begin{equation}
\eta=\sqrt{1+\frac{U_0}{q} \frac{\kl^2}{{\Delta}^2+\gamma^2}}.
\end{equation}
In contrast to the linear-coupling case \cite{rotate}, $\eta$ is
independent of frequency, maximized at ${\Delta}=0$, and can be
as large as $10^3$ or even higher for typical parameter values.
Stiffer trapping and even more efficient cooling can be expected for
small values of $q$, a key factor for the detection of quantum
fluctuation effects \cite{rotor2,rotor3,rotor4}.

\begin{figure}
\includegraphics[]{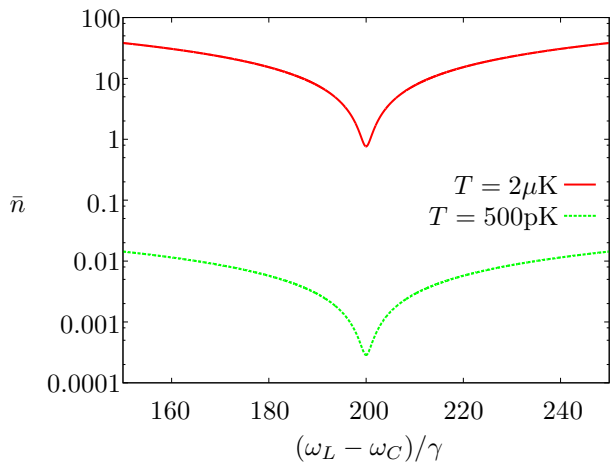}
\caption{Average roton occupation number $\bar{n}$ as a function of scaled static detuning for T= 2 $\mu$K and 500 pK, $U_0=2\pi\times 100~\mathrm{Hz}, \gamma=2\pi\times 50~\mathrm{kHz}, \kappa_L=2\pi\times 3~\mathrm{MHz}$. }
\end{figure}

The effect of this enhancement on the mean roton number can be found
from the first-order correlation functions of the system. From Eq.
(\ref{EOM}) we find
\begin{align}
\frac{d}{dt}\left<\theta^2\right>=&\frac{1}{I}\left<L_z\theta+\theta
L_z\right> \, , \nonumber\\
\frac{d}{dt}\left<L^2_z\right>=&[-I\omega^2_\theta-2\xi_\theta|a_{\rm s}|^2]\left<L_z\theta+\theta
L_z\right> \nonumber\\
&-2\frac{D_\theta}{I}\left<L_z^2\right>+2I\frac{D_\theta}{I}(n+1/2)\omega_\theta \, , \nonumber\\
\frac{d}{dt}\left<L_z \theta + \theta
L_z\right>=&\frac{2}{I}\left<L_z^2\right>+2(-I\omega_\theta^2-2\xi_\theta|a_{\rm s}|^2)\left<\theta^2\right>\nonumber\\
&-\frac{D_\theta}{I}\left<L_z \theta + \theta L_z\right> \, ,
\label{correom}
\end{align}
where the term featuring $n=[\exp(\omega_\theta/k_BT)-1]^{-1}$, the mean thermal
excitation number at temperature $T$, accounts for a shift of the first-order correlations due to contact with a thermal reservoir. The steady state solution to Eq.~\eqref{correom} is stable and exhibits non-vanishing expectations $\langle \theta_z^2 \rangle_{\rm ss}$ and $\langle L_z^2 \rangle_{\rm ss}$.
The steady-state energy of the quantized rotor is written
$E_{\rm Q} = \langle L_z^2 \rangle_{\rm ss}/2I + I \omega_{\theta}^2 \langle \theta_z^2 \rangle_{\rm ss} / 2$. Using the steady-state solution of Eq.~\eqref{correom} one finds
$E_{\rm Q} = (n+ 1/2  ) \omega'_\theta$, where $\omega'_\theta=(\omega_\theta/2) [1+\eta^{-2} ]$. Setting $E_{\rm Q} = \bar{n} \omega_{\rm eff}$ then gives
the roton occupation number
\begin{equation}
\bar{n}=\frac{(n+1/2)\omega'_\theta}{\omega_{\mathrm{eff}}}=\left (n+\frac12\right )\frac{\eta^2+1}{2\eta^3} \, .
\end{equation}
The  large trap frequency enhancement $\eta$ resulting from the COM
coupling enables one to reach the otherwise very elusive quantum
regime $\bar{n}<1$. Without that effect, a typical temperature of
$T\sim 2~\mu \mathrm{K}$ \cite{95Na} gives $n\sim 4\times 10^{3}$
for $q/c_2=10^{-3}$, indicative of large thermal excitations of the
rotor. The COM enhancement, with typical values of $U_0\sim
2\pi\times 100~\mathrm{Hz}$, $c_2\sim 2\pi\times 20~\mathrm{Hz}$
(for $^{23}$Na atoms), $N\sim 10^{5}$ \cite{optobec}, $\gamma \sim
2\pi \times 50~\mathrm{kHz}$, and $\kappa \sim 2\pi \times
3~\mathrm{MHz}$, gives $\bar{n}_{\mathrm{min}}<1$ for all suitable
values of $q$. For very low temperatures $T\sim 500~\mathrm{pK}$
\cite{pK}, we have $\bar{n}_{\mathrm{min}}\sim 10^{-4}$ (see Fig.
1), illustrating how the COM technique opens up the possibility to
observe a single-roton state \cite{1photon} or even quantum jumps of
a spin gas \cite{1photon,disper,disperbec}.

In summary, we have investigated the optomechanics of an antiferromagnetic BEC by using a formal analogy with a torsional oscillator that is quadratically coupled to a single cavity mode. The resulting bistability-free effect, the noise spectra of both the atoms and the transmitted photons, and
the quantized energy of the atoms were discussed.
That proposal provides an ideal nondestructive tool for the control
of quantum spin dynamics~\cite{rotor1,rotor2,rotor3}, and
facilitates the experimental study of the deep quantum regime of the
rotor model, which has been previously unattainable. Furthermore,
since the system we describe is within the reach of current
experimental capabilities, our results can be of immediate use to
researchers studying such diverse problems as  spin-based
nano-mechanical devices~\cite{Macrospin1}, squeezed rotors
\cite{nunnen}, hybrid spin-mirror entanglement, and high-precision
control of spinor atoms, polar molecules, or even rotating
BECs~\cite{antivor}.


This work is supported by the DARPA ORCHID program through a grant
from AFOSR, the U.S. National Science Foundation, and the U.S. Army
Research Office. H.J. thanks R. Barnett for helpful discussions.

Note added: After this manuscript was submitted, another paper~\cite{Macrospin1} presented a related theoretical analysis on a torsional model of a cavity-spin system.

\end{document}